# RISTRETTO: Manufacturing of a single-mode visible high resolution spectrograph

Bruno Chazelas[a*], Christophe Lovis[a], Nicolas Blind[a], Ludovic Genolet[a], Ian Hughes[a], Michael Sordet[a], Robin Schnell[a], Anthony Carvalho[a], Maddalena Bugatti[a], Adrien Crausaz[a], Samuel Rihs[a], Pablo Santos Diaz[a], David Ehrenreich[a], Emeline Bolmont[a], Christoph Mordasini[b], Martin Turbet[c].

[a]Observatoire de Genève, University of Geneva, 51 chemin de Pegasi 1290 Versoix, Switzerland,
[a]Physikalisches Institut, Universität Bern, Gesellschaftsstrasse 6, CH-3012 Bern, Switzerland,
[c]Laboratoire de Météorologie Dynamique, IPSL, CNRS, Sorbonne Universitée, 4 place Jussieu, F-75252 Paris Cedex 05, France

## ABSTRACT

The Spectrograph of the RISTRETTO instrument is now currently being manufactured. RISTRETTO is an instrument designed to detect and characterize the reflected light of nearby exoplanets. It combines high contrast imaging and high resolution spectroscopy to detect the light of exoplanets. The high resolution spectrograph subject of this paper uses the doppler effect to disentangle the planetary signal from the stellar light leaks. In this paper we describe the final design of the spectrograph and report the status of its construction. The RISTRETTO spectrograph has seven diffraction limited spaxels. The spectrograph's resolution is 130000 in the 620-840 nm band. It is designed in a similar way as HARPS and ESPRESSO, being a warm, thermally controlled spectrograph under vacuum. It is designed to be compact and self contained so that it could be installed on different telescopes. It is however tailored to be installed on a nasmyth platform of a VLT telescope. We present updates to the design and the manufacturing of the instrument. In particular we present the performance of the thermal enclosure.

## 1. INTRODUCTION

RISTRETTO[1–3] is a 2-stage instrument combining extreme adaptive optics[4–7], coronography[8] and high resolution spectroscopy. The spectrograph is the subject of this paper. It is a single mode high resolution spectrograph with seven spaxels, fed with an IFU that will be connected to the front -end part of the instrument. The spectrograph is designed to operate from 620 nm to 840 nm with a resolution greater than 130000.

The spectrograph is designed as a double-pass Echelle spectrograph. Due to the very small etendue of the single mode fibers, the whole instrument fits in a vacuum tank of 85 cm in diameter. To keep it stable at the level of 10 m.s$^{-1}$, the spectrograph is kept under vacuum (< 10$^{-4}$ mBar) and thermally controlled to better than 20 mK.

To accomodate all the diffraction orders for the different fibers, a 4K detector is required. A deeply depleted Teledyne E2V CCD231-84-x-E74 detector with an astra-multi-2 coating was chosen.

The entire spectrograph is housed in a thermally insulated enclosure that actively maintains a constant temperature while keeping the thermal load to the environment as lower than the standard VLT ESO requirement for a Nasmyth instrument. The Instrument is mounted on a steel frame and is easy to transport as is. Figure 1 shows the view of the spectrograph and its control system, installed on a VLT Nasmyth platform. The full instrument shall occupy a full Nasmyth platform.

There is still enough space (less than half the area is occupied by the spectrograph), and weight margin (the spectrograph wight is 2.5 T out of the 8T available on the Nasmyth platform of a VLT[9].) on the Nasmyth platform for the front-end.

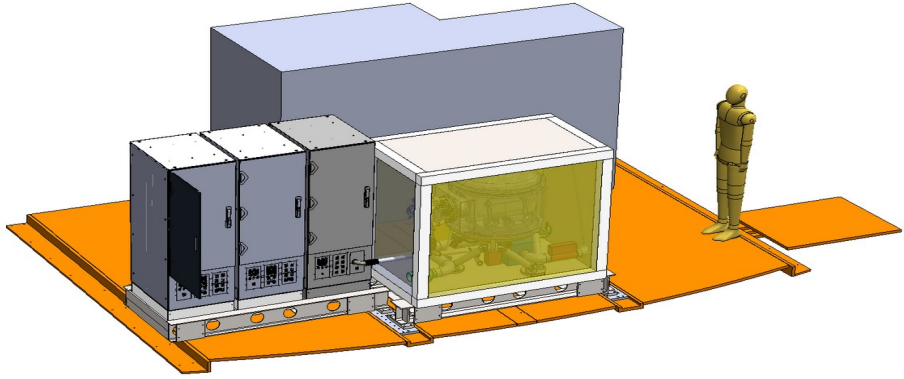

Figure 1: Layout of RISTRETTO on the Nasmyth platform. There is space and weight allocation for the XAO front end. The entire Nasmyth platform is needed.

## 2. SPECTRAL FORMAT

The spectral format of the spectrograph is shown on figure 2. The resolution of the instrument is limited by diffraction and actual optical quality. Thanks to the PyEchelle simulator, it is possible to simulate full detector frames and start learning how to use RISTRETTO spectra to detect exoplanets[10].

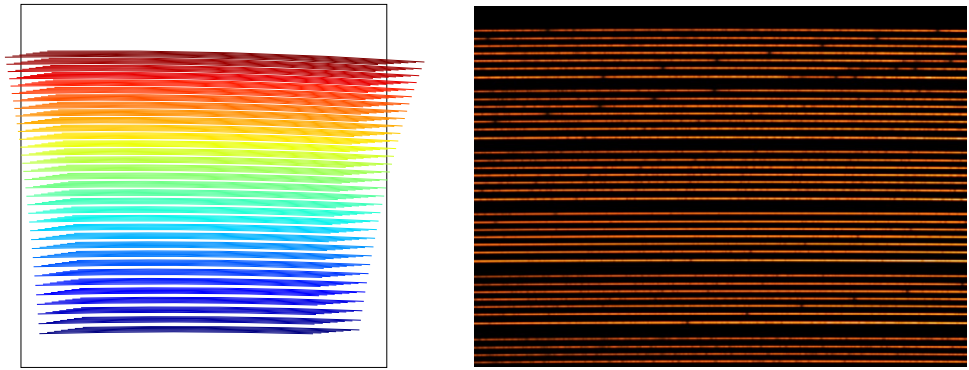

Figure 2: Spectral format of the spectrograph. There are 34 orders, each having seven fibers. The seven fibers are on a nearly horizontal slit, in order to get the required separation for the different fibers. On the right, a simulation with PyEchelle[13] of the spectrograph. Here a stellar spectrum. Some spectral features are visible and show how the slit is actually tilted.

## 3. OPTICAL MANUFACTURING

As the resolution of the spectrograph is limited by the diffraction, great care has been taken to manufacture lenses of high optical quality ($\lambda/20$ rms in average). For the last lens of the spectrograph, a field lens that serves as a field lens and at the same time is the detector cryostat window, the optical quality *per se* is not a good criterion since a individual beam at any given wavelength has a very small impact area on the lens, instead, in order to avoid distortion of the diffraction order the choice was made to ask for a traditional polishing technique that provides a nice 1/f spatial frequency decrease in the defects.

We are still waiting for the cross-disperser prism and for the fabrication of a test fiber link.

The grating chosen is an off-the-shelf grating from MKS/Grating Labs. It should be the worse optical component in terms of optical quality. It provides a $\lambda/4$ PV WFE.

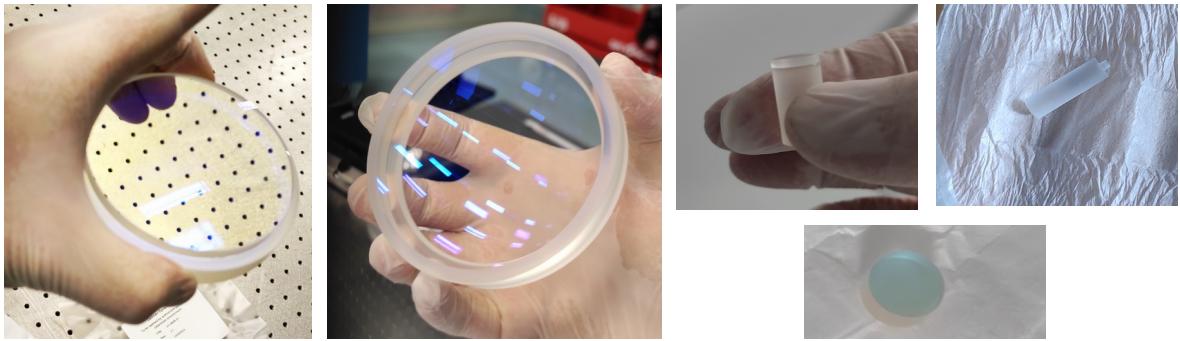

Figure 3: Photo of the different lenses of the spectrograph that have been already manufactured.

## 4. THERMAL ENCLOSURE

The thermal enclosure is designed to keep the spectrograph at a constant temperature while not heating the dome atmosphere to create as little turbulence as possible.

The main challenge is that we need to use a target temperature for the spectrograph that is below the maximum temperature of the dome otherwise we would need a too much insulation and too much cooling power. Analytical calculations and then an OpenModelica model helped to find the right temperature target and the right dimensioning for our system. The conceptual scheme of the thermal control system is shown on figure 4. A water-glycol circuit runs between the control rack and the thermal enclosure, at a constant temperature, thanks to a Peltier-effect based chiller (200W of cooling capacity / 600W of heating capacity). This makes it possible to either cool or heat the water, managing the different situations with low vibrations. The excess heat is dissipated using the control cabinet cooling system.The thermal enclosure contains a water/air heat exchanger and a forced air circulation that blows air over the vacuum tank. In the air circulation there are heaters (100W) that are used to regulate the temperature. Figure 5 shows the thermal enclosure before closing it. The insulating box is made out of PUR with 0,8 mm steel skins to have an efficient, light and structural material.

In addition, to avoid thermal disturbances through the support structure of the tank, the feet of the tank located in the thermal enclosure are thermally insulated from the frame of the spectrograph with G10 sheet and thermally regulated with flexible polymide heaters. This solution has proved to be effective particularly for the seasonal thermal variations on HARPS and ESPRESSO.

The performances of the enclosure are excellent overall. We tested it outside of our workshop at the Geneva Observatory throughout the end of winter and spring in conditions much more challenging than those at the VLT's nasymth. After some adjustments to the thermal loops and the different temperature setpoints, we achieved the performances shown on the figure 6. The temperature of the bottom of the tank, used to regulate the temperature of the enclosure, achieves a stability better than ± 0.05K. So the heat load disturbances are minimized at the mechanical interface between the tank and the optical bench leading to a much stable heat transfer by conduction. The upper parts of the tank have a stability of ±0.5K which is better than the requirement of ±1K. A relaxed stability is acceptable because the heat is less efficiently transferred by radiation between the upper parts of the tank and the optical bench. Moreover, a thermally controlled radiation shields further minimize the heat load disturbances by insulating the optical bench from the tank..

The system was designed to comply with the ESO requirement[9] for thermal load on the Nasmyth platform of a VLT:
- no more that 150 W for the heat load to the outside air,
- +/- 1.5°C between skin temperature of the instrument and ambient air (for a wind speed of 2m/s and a convective heat transfer coefficient of of h = 2W/m2/K).

The latter requirement is based on an assumption about the convection coefficient, which makes is difficult to verify directly measuring the surface temperature as we cannot put ourselves in the condition stated in the requirement. The method we have devised involves measuring the output power from the enclosure using the temperature difference between inside and outside and the known thermal resistance for the insulation panel we are using. This is then compared to the power that would be dissipated in the specified conditions to give us a limit.

The power limit is $Q = h.\Delta T.A_{enclosure} = 45W$. By applying this concept we have produced figure 7 which demonstrates that we are compliant with the requirement.

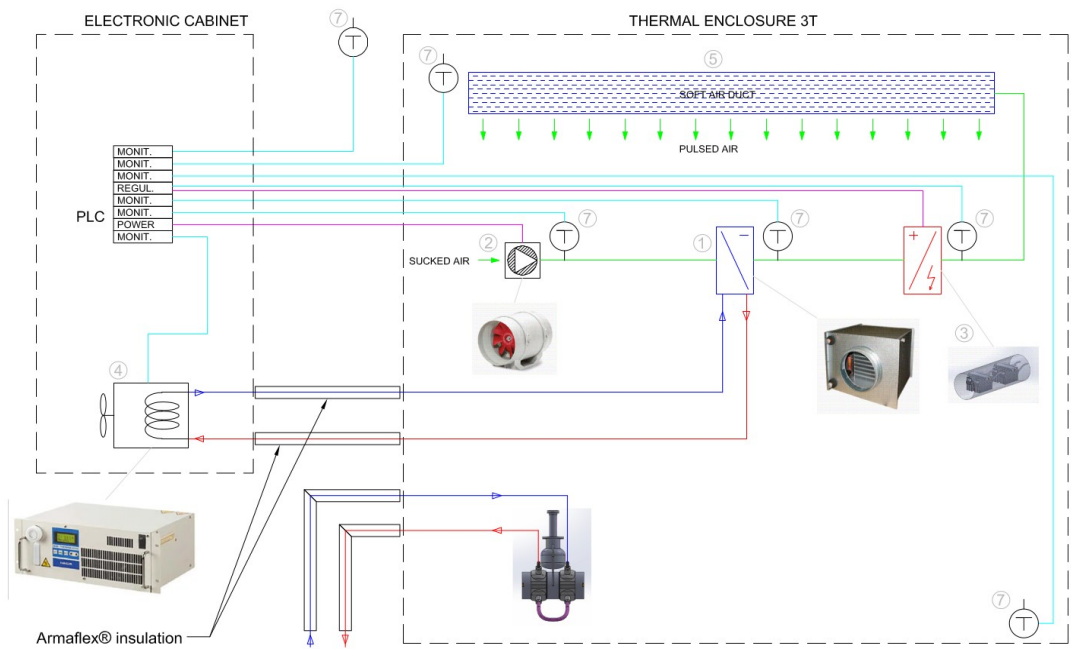

Figure 4: Scheme of the thermal enclosure first layer of control. An air cooled chiller with a Peltier system provides a stable water temperature to an air water exchanger in the thermal enclosure. A forced air circuit passes through the exchanger and through heaters that regulate the air temperature. This stage provides a thermal stability of +/- 1° to the vacuum tank.

Unlike previous similar developments for the thermal control of Calibration Fabry-Perot Cavities or spectrograph such as HARPS, where we have used integrated controllers from Lakeshore, for this enclosure we have used a combination of Lakeshore 240 temperature input modules, which can be easily connected to a PLC via a PROFIBUS interface. The control loop then is implemented using a Beckoff PLC. The PLC activate a solid state relay that modulates the heating power in the enclosure in a PWM scheme (a 1-second cycle with a 10 ms resolution). This has proven to be an effective solution. Figure 8 illustrates the scheme. The solution is effective for objects with a slow thermal response to with respect to the PWM cycle.

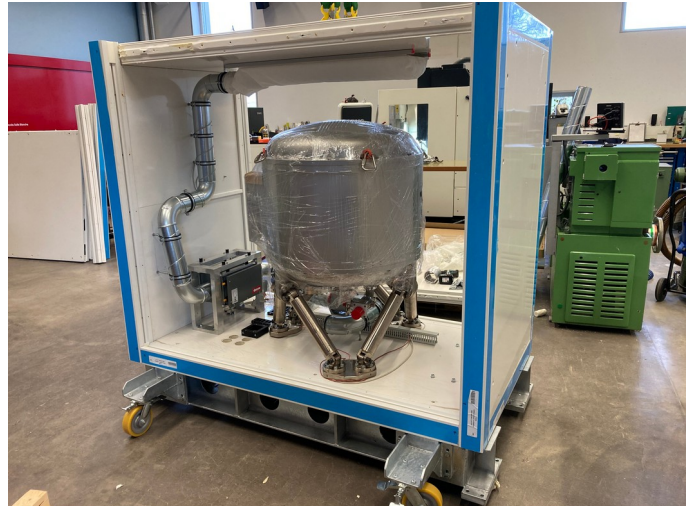

Figure 5: Photo of the thermal enclosure before closing. One also can see the vacuum tank.

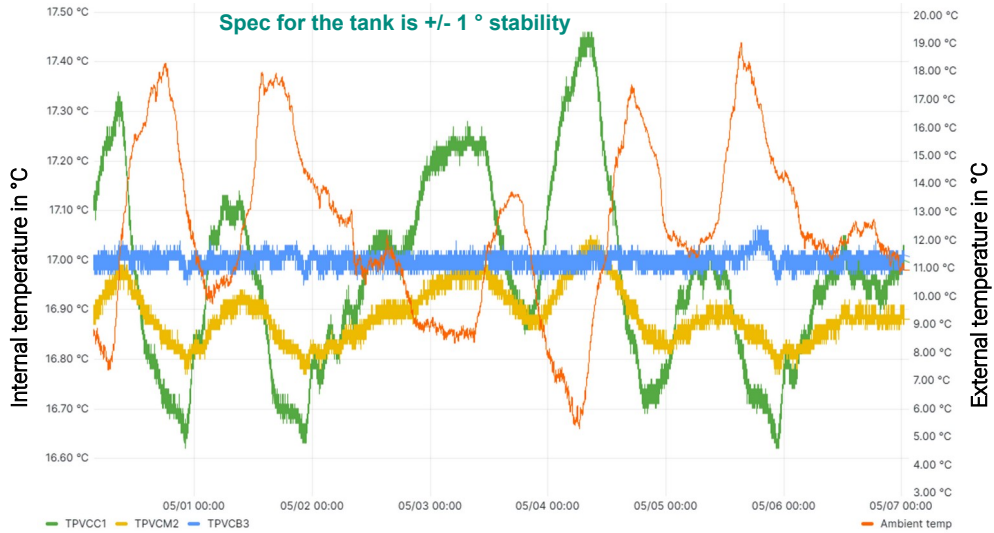

Figure 6: Performance of the first layer of the thermal enclosure, while sitting outside the Geneva Observatory during spring 2024.

**Red** is Ambient temperature outside
**Blue** and **Yellow** bottom of the tank where optical bench will be fixed
(temperature measured outside the tank)

**Green** is the top temperature of the tank, it is currently directly under the air flow, maybe could be slightly shielded

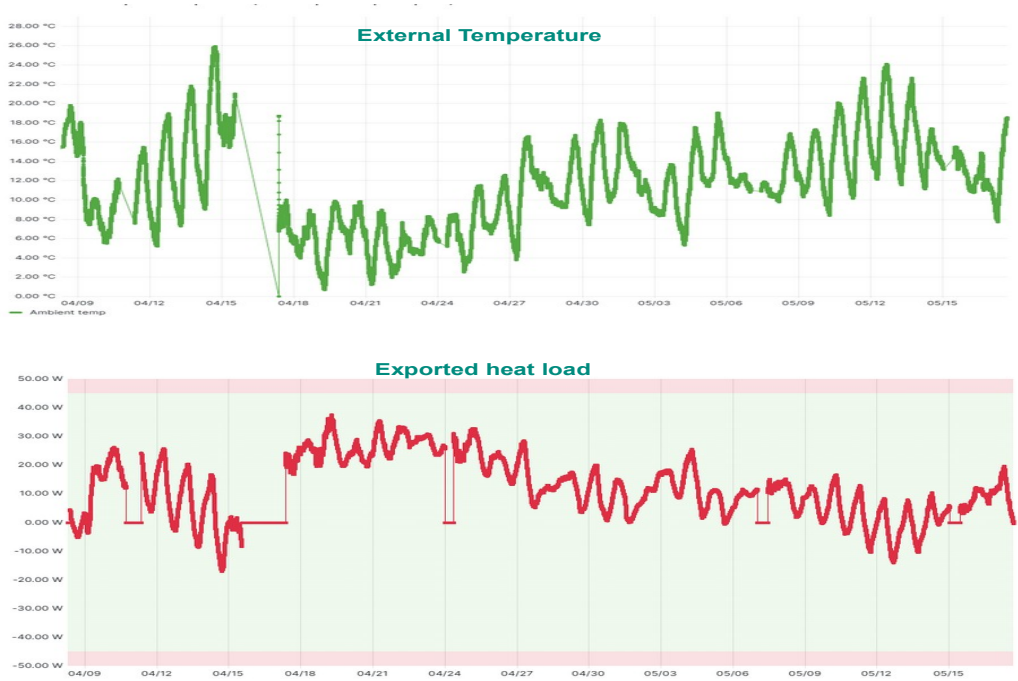

Figure 7: Thermal load to the environment, computed using the temperature difference between the inside and the outside of the thermal enclosure and the insulation panels thermal resistance. Data taken in spring 2024, with the Thermal enclosure sitting outside of the Geneva Observatory. The system always stayed in the allowed envelope.

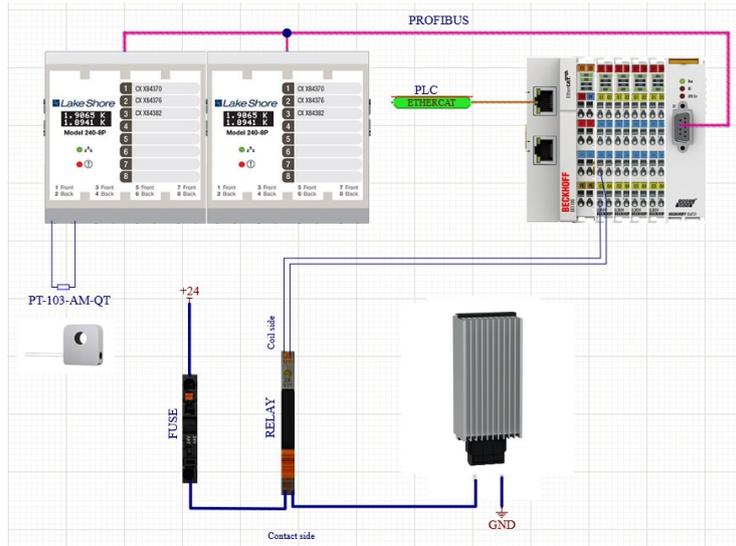

Figure 8: Concept of one thermal loop in the thermal control system. Thermal sensors are read using Lakeshore 240, PLC friendly, temperature input modules. The heaters power is modulated using PWM signal triggering a solid state relay.

## 5. DETECTOR HEAD

The RISTRETTO spectrograph is an EPRV instrument under warm vacuum. The detector head is thus a differential vacuum cryostat where the detector is attached to the optical table rigidly, while being softly connected to the tank through a flexible bellow. It is cooled by a LPT9310 cryocooler from Thales equipped with an active vibration reduction system. Special care has been taken to reduce the vibrations generated by the cryocooler and transmitted to the instrument and the rest of the telescope.

The manufacturing of the different detector head parts is nearly complete. Figure 9 shows a subset of the different produced parts. The next step is pressure testing, before the first cryogenic test with a thermal equivalent of the detectors. Once these tests will have been completed, the real detector will be integrated and the whole system will be characterised. This should happen before the end of 2024.

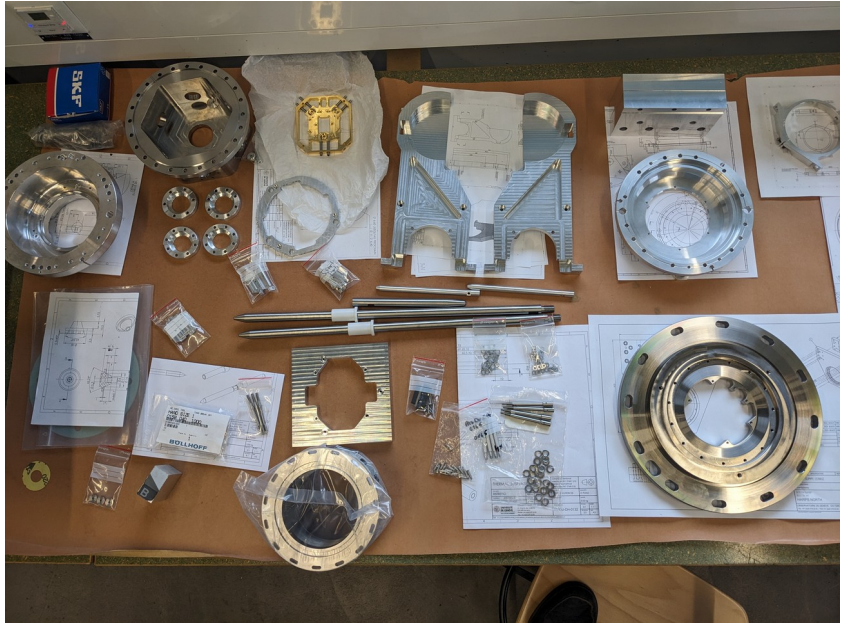

Figure 9: parts produced already for the detector head

## 6. CONCLUSION

This paper presents the current status of the construction of the RISTRETTO spectrograph. Subsystems are being tested as soon as they are ready. The performances of the thermal enclosure have been shown to be in line with our internal requirement and respect ESO VLT thermal requirements. The MAIV of the spectrograph is scheduled for completion by mid- 2025. Once it is ready the instrument will undergo laboratory tests and then on-sky testing. Two possible location for on-sky testing are considered the PAPYRUS[11] AO at OHP and/or on the EULER telescope paired with the KalAO[12] adaptive optic system.

## 7. ACKNOWLEDGMENTS

We would like to thanks in particular Bernard Delabre that helped us with the spectrograph design concept.

This work has been carried out within the framework of the National Centre of Competence in Research PlanetS supported by the Swiss National Science Foundation under grants 51NF40_182901 and 51NF40_205606. The RISTRETTO project was partially funded through the SNSF FLARE programme for large infrastructures under grants 20FL21_173604 and 20FL20_186177. The authors acknowledge the financial support of the SNSF.

Huge thanks to Julian Stuermer for help with using PyEchelle[13] (https://gitlab.com/Stuermer/pyechelle)!